# Two level anti-crossings high up in the single-particle energy spectrum of a quantum dot

C Payette,[1,2,*] D G Austing,[1,2] G Yu,[1] J A Gupta,[1]
S V Nair,[3] B Partoens,[4] S Amaha,[5] S Tarucha[5,6]

[1]*Institute for Microstructural Sciences M50, National Research Council of Canada, 1200 Montreal Road, Ottawa, Ontario K1A 0R6, Canada*

[2]*McGill University, Department of Physics, Ernest Rutherford Physics Building, 3600 rue University, Montréal, Quebec, H3A 2T8, Canada*

[3]*University of Toronto, Center for Nanotechnology, 170 College St., Toronto, Ontario, M5S 3E4, Canada*

[4]*Departement Fysica, Universiteit Antwerpen, Groenenborgerlaan 171, B-2020 Antwerpen, Belgium*

[5]*ICORP Quantum Spin Project, JST, 3-1, Morinosato-Wakamiya, Atsugi, Kanagawa, Japan*

[6]*Department of Applied Physics, University of Tokyo, 7-3-1, Hongo, Bunkyo-ku, Tokyo, Japan*

**Abstract**

We study the evolution with magnetic field of the single-particle energy levels high up in the energy spectrum of one dot as probed by the ground state of the adjacent dot in a weakly coupled vertical quantum dot molecule. We find that the observed spectrum is generally well accounted for by the calculated spectrum for a two-dimensional elliptical parabolic confining potential, *except* in several regions where two or more single-particle levels approach each other. We focus on two two-level crossing regions which show unexpected anti-crossing behavior and contrasting current dependences. Within a simple coherent level mixing picture, we can model the current carried through the coupled states of the probed dot provided the intrinsic variation with magnetic field of the current through the states (as if they were uncoupled) is accounted for by an appropriate interpolation scheme.

*PACS:* 73.21.-b; 73.22.Dj; 73.21.La; 73.23.Hk

Keywords: Quantum Dots; Quantum Dot Molecules; Fock-Darwin Spectrum; Anti-crossings; Single Electron Resonant Tunneling

## 1. Introduction

Energy level mixing in coupled low dimensional systems is a basic quantum phenomena leading to level anti-crossing and superposition effects. Some examples are: wavefunction and level mixing in spatially-coincident and spatially-separated coupled one-dimensional electron systems [1-3]; anti-crossing of excitonic transitions in self-assembled quantum dot molecules [4,5]; anti-crossing of phonon sidebands and exciton states in nanocrystals [6]; self-assembled quantum dot exciton-photonic crystal cavity mode coupling [7]; and avoided level crossings of states in few-electron (N>1) single and coupled quantum dots in the presence of Coulomb interactions [8].

Weakly coupled vertical quantum dots (QDs) with *nearly ideal but not perfect* lateral confinement potentials are another coupled low dimensional system to exhibit strong and non-trivial level mixing behavior as we outline below. The level mixing is between single-particle states within each QD, i.e. intra-dot rather than inter-dot, and the levels are brought into proximity by applying a magnetic field.

---
[*] Corresponding author. Tel.:+1-613-990-7019; fax: +1-613-990-0202; e-mail: chris.payette@nrc-cnrc.gc.ca

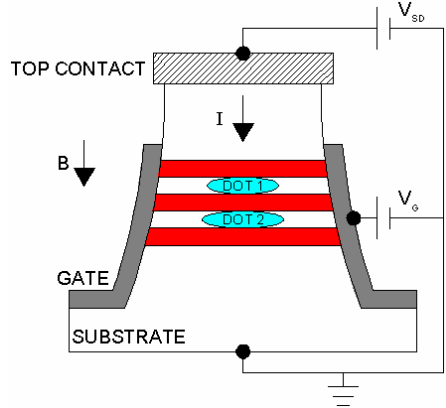

Fig 1: Schematic diagram of a gated vertical quantum dot molecule device made from a triple-barrier, double-well resonant tunneling structure.

## 2. Device Structure and Measurement Principle

Our sub-micron circular mesa device, shown schematically in Fig. 1, is fabricated from a structure containing three $Al_{0.22}Ga_{0.78}As$ tunnel barriers and two $In_{0.05}Ga_{0.95}As$ quantum wells. The quantum dot regions are thus strongly confined in the growth direction by hetero-structure barriers. Weaker lateral confinement of the electrons in the dots is achieved by applying a voltage to the Schottky side gate wrapped around the base of the mesa to modulate electrostatic depletion from the mesa side wall. The two quantum dots are weakly coupled in the vertical direction with a tunnel coupling energy ($\Delta_{SAS}$) of less than 0.1 meV. Current (I) flows through the two dots in response to an applied bias between the top and substrate contacts. Furthermore, we apply a magnetic field (B) in a direction parallel to the direction of the current. Measurements are performed at ~0.3 K.

To measure the spectra of the constituent dots in the double dot structure, we employ the following technique. By simultaneously changing the bias voltage between the top contact and the grounded substrate (source-drain voltage, $V_{SD}$) and the voltage on the gate, $V_G$, we can arrange for the lowest energy (1s-like) state in the upstream QD to resonantly probe the Fock-Darwin-like spectrum of the downstream QD in the *single- electron tunneling regime*. This measurement technique is more fully described in Ref. [9].

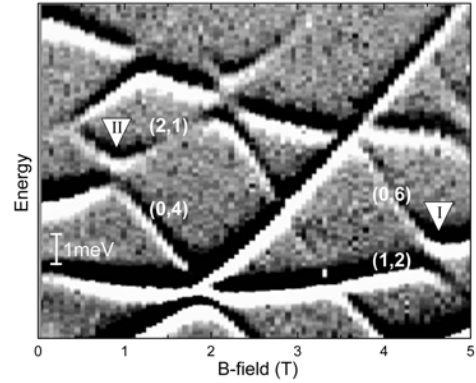

Fig 2: Grey scale plot of experimental differential conductance ($dI/dV_{SD}$) data. The single-particle levels of the probed dot are mapped out by the black-white stripes. The vertical axis corresponds to energy and the white scale bar represents ~1 meV. The energy scale is deduced by comparing the magnetic field positions of the measured level crossings with the same level crossings in the calculated single-particle spectrum. Strong variation in the appearance of the stripes with magnetic field is evident.

## 3. Discussion

Assuming that the lateral confining potential in our QDs is *circular or elliptical and parabolic*, we would expect the single-particle levels of the probed dot to evolve with B in a very distinct and predictable manner [10,11], and all level crossings should be 'exact' crossings. Any deviations from this idealistic potential, however, may cause levels to anti-cross. Thus, the precise details of observed anti-crossings in dot spectra can in principle shed light on the nature of deviations in confining potentials of realistic dots.

As a typical example, one of the dots in the device we have studied is known to have a single-particle spectrum which overall can be fitted very well to the single-particle spectrum for an elliptical parabolic dot with ellipticity ~4/3, and major and minor axes confinement energies respectively of ~4.6 and 6.1 meV [12]. However, the part of the experimental spectrum shown in Fig. 2 displays clear and unexpected anti-crossing behavior, with levels split by several hundreds of µeV, at many of the level crossings. In this paper, we will focus on the two-level anti-crossings identified in Fig. 2 as I and II. The particularly interesting three-level crossing behavior will be discussed in detail elsewhere [13].

The precise microscopic origin of the observed anti-crossings is unknown. Picturing deviations from an ideal elliptical and parabolic lateral confining potential as perturbations, we assume they are due to higher order terms (beyond terms like $x^2$ and $y^2$), caused by anharmonicity and anisotropy, present in the lateral confining potential of the probed dot. We stress that even in the vertical geometry real QDs are never perfectly circular or elliptical and parabolic.

Despite not knowing the exact form of the lateral confining potential with the appropriate higher order terms, we can none the less study the anti-crossing behavior in the spectrum of the probed dot. Examining regions, like I and II in Fig. 2, where two levels anti-cross, we first deduce the coupling energy by fitting the position of the black-white conductance stripes which map the upper and lower current resonance branches. We assume that the two states have a linear energy dispersion with magnetic field in the vicinity of the crossing if they were uncoupled. We then compute the current carried through the coupled states assuming the level mixing within the probed dot is coherent. Full details of our model will be published elsewhere [13].

Examining Fig. 2 in more detail, since, overall, we can explain the experimental spectrum well with the spectrum of an elliptical quantum dot, we can still label the states with the quantum numbers $(n_x, n_y)$ appropriate for parabolic confinement [11]. The anti-crossings of interest are thus between the $(n_x, n_y) = (0,6)$ and $(1,2)$ states (labeled I), and between the $(n_x, n_y) = (2,1)$ and $(0,4)$ states (labeled II). The coupling energy for the states at I and II is approximately 0.5 and 0.4 meV respectively.

Considering first the resonant current dependence for anti-crossing I, the crossing has a distinct upper branch and a distinct lower branch as shown in Fig. 2. We plot in Fig. 3, the resonant current through each of these branches, where we have subtracted out the non-resonant background current. We see that to the left of the crossing region the lower (upper) branch is strong (weak), while to the right the lower (upper) branch is weak (strong). This behavior is what one might anticipate as 'typical' anti-crossing behavior, and has been seen, for instance, in the relative photoluminescence intensity for two anti-crossing excitons [4].

Turning now to the anti-crossing II we see again a distinct upper and lower branch in Fig. 2. However, the dependence of the resonant current is markedly different from that of anti-crossing I. Referring to Fig. 4, to the left of the crossing, the lower (upper) branch is strong (weak). On passing through the crossing, rather than have the branch strengths rapidly interchange as in anti-crossing I we can see that the two branches have roughly equal strengths in the region between 0.8 and 1.2 T. To the right of the crossing, the lower branch begins to recover its strength, whilst the upper branch becomes weak again. This behavior on first sight is unanticipated anti-crossing behavior and differs from that observed for two anti-crossing excitons in Ref. [4] for example.

Our model uses a least squares technique to simultaneously fit both the measured current through, and the position of, each of the resonances. Naively, one might have assumed that the resonant current through the constituent states *even in the absence of the coupling* is independent of magnetic field. However, if we make no attempt to account for the underlying variation of the current carried by the probed dot's states with magnetic field, the resulting fits shown as thin lines in Figs. 3 and 4 are inadequate. For anti-crossing I, the fit matches the basic shape of the measured current, but fails to model the current quantitatively. For anti-crossing II, the fit is quite poor, failing to show even qualitatively the correct dependence.

In order to improve the fit, we can include in our model the underlying variation with magnetic field of the resonance current through the states in the vicinity of the crossing. This variation arises due to the non-ideal nature of the confining potential which influences the strengths of the resonances throughout the entire spectrum. The substantially improved fits of the resonance current are shown as thick lines in Figs. 3 and 4. A challenge arose as to how to interpolate the resonance current from the low field side to the high field side of the crossing. We have experimented with two interpolation schemes: one based on linear interpolation of the measured current itself and the other based on linear interpolation of the current *amplitude* ("square root" of the current which in our scheme may be positive or negative).

We find that in some cases, such as for anti-crossing I, the *amplitude* interpolation scheme produces the best result, whilst for other cases, such as anti-crossing II, the current interpolation scheme works best. Further study is required to determine why different interpolation schemes are needed [13].

**Conclusion**

We have studied different examples of the two level anti-crossing behavior observed in the single-particle energy spectrum of a single dot in a weakly coupled vertical quantum dot molecule. Our simple model which computes the current in a coherent level mixing picture allows us to fit the observed behavior of the resonances well when we account for the underlying variation with the magnetic field of the resonance current in the vicinity of the anti-crossing.

**Acknowledgements**

We are grateful for the assistance of A. Bezinger, D. Roth, and M. Malloy for micro-fabrication. CP is funded by DGAs NSERC Discovery Grant. Part of this work is financially supported by the Grant-in-Aid for Scientific Research S (No. 19104007), SORST-JST, Special Coordination Funds for Promoting Science and Technology, MEXT, and the DARPA-QUIST program (DAAD19-01-1-0659).

**References**

[1] K. J. Thomas, et al., Phys. Rev. B 59 (1999) 12252.
[2] G. Salis, et al., Phys. Rev. B 60 (1999) 7756.
[3] S. F. Fischer, et al., Nature Physics 2 (2006) 91.
[4] H. J. Krenner, et al., Phys. Rev. Lett. 94 (2005) 057402.
[5] E. A. Stinaff, et al., Science 311 (2006) 636.
[6] L. Zimin, et al., Phys. Rev. Lett. 80 (1998) 3105.
[7] K. Hennessy, et al., Nature 445 (2007) 896.
[8] See recent review article by R. Hanson, et al., cond-mat/0610433v1.
[9] K. Ono, et al., Physica B 314 (2002) 450.
[10] V. Fock, Z. Phys. 46 (1928) 446; C.G. Darwin, Proc. Cambridge Philos. Soc. 27 (1930) 86.
[11] A. V. Madhav and T. Chakraborty, Phys. Rev. B 49 (1994) 8163.
[12] D. G. Austing, et al., Phys. Stat. Sol. (a) 204 (2007) 508.
[13] C. Payette, et al., (2007) unpublished.

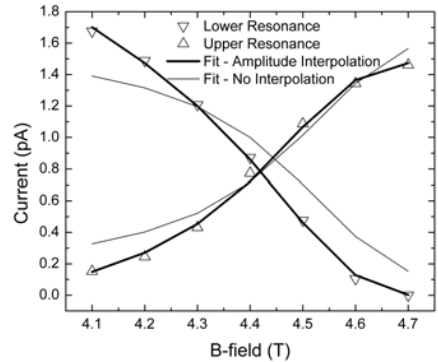

Fig 3: Measured resonant current through the lower branch (▽) and upper branch (△) resonances with background current subtracted for anti-crossing I. The thin (thick) line is a fit of the data without any interpolation (with amplitude interpolation).

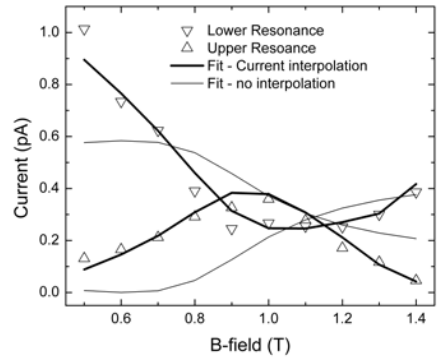

Fig 4: Measured resonant current through the lower branch (▽) and upper branch (△) resonances with background current subtracted for anti-crossing II. The thin (thick) line is a fit of the data without any interpolation (with current interpolation).